\documentclass[a4paper]{article}
\usepackage[margin=25mm]{geometry}
\usepackage{algorithmic}
\usepackage[linesnumbered,ruled,vlined]{algorithm2e}
\usepackage{amsmath}
\usepackage{amsfonts}
\usepackage{amssymb}
\usepackage{graphicx}
\usepackage{verbatim}
\immediate\write18{texcount -tex -sum  \jobname.tex > \jobname.wordcount.tex}

\providecommand{\keywords}[1]
{
  \small	
  \textbf{\textit{Keywords---}} #1
}

\title{Cross-chain between a Parent Chain and Multiple Side Chains}
\author{Guangsheng Yu$^{1}$, Xu Wang$^{2}$, Ren Ping Liu$^{2}$  \\
        \small $^{1}$Data61 CSIRO, Sydney, Australia \\
        \small $^{2}$Global Big Data Technologies Centre, University of Technology Sydney, Australia \\
}
\date{} 

\begin{document}
\maketitle

\begin{abstract}
In certain Blockchain systems, multiple Blockchains are required to operate cooperatively for security, performance, and capacity considerations. This invention defines a cross-chain mechanism where a main Blockchain issues the tokens, which can then be transferred and used in multiple side Blockchains to drive their operations. A set of witnesses are created to securely manage the token exchange across the main chain and multiple side chains. The system decouples the consensus algorithms between the main chain and side chains. We also discuss the coexistence of the main tokens and the native tokens in the side chains.
\end{abstract} \hspace{10pt}

\keywords{blockchain, cross-chain, Internet-of-Things}

\section{Introduction}
Blockchain, featured with its decentralized tamper-resistance, has been an active research topic to address the security issue of centralization, and has been widely adopted in many scenarios such as Internet-of-Things (IoT)~\cite{iot-1,iot-2,iot-3,iot-4,iot-5,iot-6} in order to achieve distributed trust~\cite{2023arXiv230608056K}. A public Blockchain, often referred to as \textit{Token Chain}, can accomplish a decentralized tokenomics. Since \textit{Token Chains} are usually public, they require decentralized consensus algorithms, such as Proof-of-X (PoX)~\cite{nakamoto2008bitcoin, YU2020101934}. 
Additionally, blockchain governance plays a crucial role in the effective management and operation of decentralized systems, guiding the decision-making processes within the network~\cite{10092670,10.1145/3551902.3564802,LIU2022102090,liu2022bgra}.
However, a decentralized consensus algorithm entails the comparatively low performance and efficiency, implying that those delay-sensitive applications or applications with massive deployment and data, such as IoT applications, cannot operate on \textit{Token Chains} properly. For such applications, consortium chains are proposed with partially decentralized and more efficient consensus algorithms, such as Byzantine Faulty Tolerance (BFT)~\cite{castro1999practical}.
It is important to note that blockchain scalability solutions such as sharding, which operates at layer-1, differ from cross-chain techniques, a layer-2 solution. 
While sharding divides the network into smaller parts to increase throughput, cross-chain solutions enable communication between different blockchains, providing additional flexibility and scalability~\cite{8954616,Yu2020,zixu}.
However, \textit{Side Chains} have no economic value, which prevents the trading of IoT data.

We propose a cross-chain architecture consisting of a public \textit{Token Chain} and multiple associated consortium \textit{Side Chains}. In the cross-chain system, the \textit{Token Chain} issues the main tokens and maintains the official token ledger for the cryptocurrency ecosystem, while a number of \textit{Side Chains} import the tokens from the \textit{Token Chain} for the management of the massive records and transactions for applications, such as IoT. The key challenge is the design of cross-chain interactions, including token exchange between the \textit{Token Chain} and its associated \textit{Side Chains}. A set of nodes are defined as \textit{Witnesses} in the genesis block of each \textit{Side Chain}. These \textit{Witnesses} are responsible for the exchange of tokens between the \textit{Token Chain} and the \textit{Side Chains}. \\
The cross-chain system provides the flexibility to meet various capacity demands of heterogeneous applications, including but not limited to IoT applications. It achieves modularity in that the \textit{Side Chains} are decoupled from the \textit{Token Chain} through the design of \textit{Witnesses}. This system is not limited to Ethereum platform, and has no limitations on the number of \textit{IoT Chains}.
\footnote{The theoretical maximum number of IoT chains is $16^{40}\approx 1.46\times 10^{48}$ based on the 20-byte address. It is a quite large number to exceed in practice.}

\section{Related Works}
There exist several cross-chain technologies, mainly classified into Relay, Hash-Lock and Notaries.
\begin{itemize}
\item Relay: Chain $A$ can prove the facts and behaviors on Chain $B$. Technologies using Relay are including but not limited to, the following solutions:
\begin{itemize}
\item BTC-Relay~\cite{hash-lock}. It is a technology that can inter-operate between Bitcoin and Ethereum. A light client that introduces Simple Payment Verfication (SPV) is implemented by a smart contract in Ethereum so that facts on Bitcoin can be proved in Ethereum. However, it also implies that Ethereum needs to store all headers of Bitcoin, which is quite expensive.
\item Non-Interactive Proof of PoW~\cite{nipopow}. Combined with Monte Carlo method and Bisection method, this proof tries to calculate the probability that a chosen puzzle with a high difficulty falls into a specific interval. As such the PoW on Chain $B$ can be proved by including this proof of PoW on Chain $A$. However, currently this method is not general and strictly dependent from PoW consensus.
\end{itemize}
\item Hash-Lock~\cite{hash-lock}. It acts as a personal trading platform. Fair and secure assets exchange can be accomplished with the aid of a time-out mechanism. However, it does not support assets portability and it is unfriendly to the interoperation between a public chain and a consortium chain because of the possible different token-scheme on this consortium chain.
\item Notaries~\cite{notary}. It is also known as \textit{Witnesses}. \textbf{Our invention belongs to this type.} In words, there exist some nodes which run clients on both Chain $A$ and $B$. Messages (Assets) exchanged is conducted through these nodes with a necessary consensus process. Notaries is the most frequently used technology so far. There are a number of Blockchain projects announcing their schemes, including but not limited to,
\begin{itemize}
\item Hcash. It is a typical design of Notaries-liked cross-chain technology. It collectively features most of the advantages of Notaries from other similar projects. We believe that comparing our invention with Hcash will be sufficiently significant to differential our invention with other similar designs. In our opinion, notaries act as a set of nodes running on both chains, as if multiple personal trading platforms exist with a necessary consensus process. The On/Off-chain multi-signature wallet takes critical responsibilities when assets exchanging. However, this design does require users to have assets prior to the process of exchanging, which implies that similar to Hash-Lock, it does not support consortium chain very well either. Hcash does not provide the solutions to address the implications of the native gas-oriented account-balance on the consortium chain, and to generate exchangeable token for a consortium chain with the native gas-oriented account-balance.
\item Walton Chain. Notaries are a set of nodes only running on the local chain. The request of generating a new side chain is executed by a smart contract by sending a transaction. In other words, notaries are the bridge to exchange the messages of the most external layer to the internal layer with a necessary consensus process. It implies that a side chain is only an abstractive structure in Walton Chain. Link structure is maintained on smart contracts in which the data storage is simplified as the state-oriented data to be stored in the local database. Such a design is not only strongly restricted by the throughput of the most external chain but also extremely increases the loading of the entire network.
\end{itemize}
\end{itemize}
\emph{There appears an existing national patent (CN 106447309 A) in China that is similar to our invention. Their patent implements the third-party chain that pegs to both the main chain and the side chain with a certain address on each chain as if an accessible API connects the main chain and the side chain. It implies that these APIs have an offline information relaying between the certain address on either the main chain or the side chain, and its third-party chain, which they did not precisely point out. In addition, the existence of the third-party chain exposes the issue of data privacy. It also makes no sense that the third-party is eligible to maintain multiple side chains due to the data privacy, centralization and scalability.} Therefore, we can conclude that our invention is more general and outplays their patent.

Our invention is targeting at such a case that multiple \textit{IoT Chains} with different consensus algorithms, a high throughput and a high volume are anchored with a \textit{Token Chain}. Therein each \textit{IoT Chain} has its own business groups charged with the audition of cross-chain process. One \textit{IoT Chain} can be totally a black-box for the others. Likewise, any behaviors on an \textit{IoT Chain} can be also a black-box for the \textit{Token Chain} apart from the amount of token exchanged. Either an \textit{IoT Chain} without the native gas or an \textit{IoT Chain} withe the native gas can be registered to the \textit{Token Chain} whenever a user requires token involved and afterwards provides the sufficient assets and proofs on the \textit{Token Chain}. In words, we believe that our invention can be a better solution to the separation between the financial-layer and data-layer in order for a more data-friendly and healthy Blockchain system.
\section{Brief Description of the Drawings}
The description of each step and module are shown in the following drawings.

\begin{figure}[htpb]
\centering
\includegraphics[width=5in]{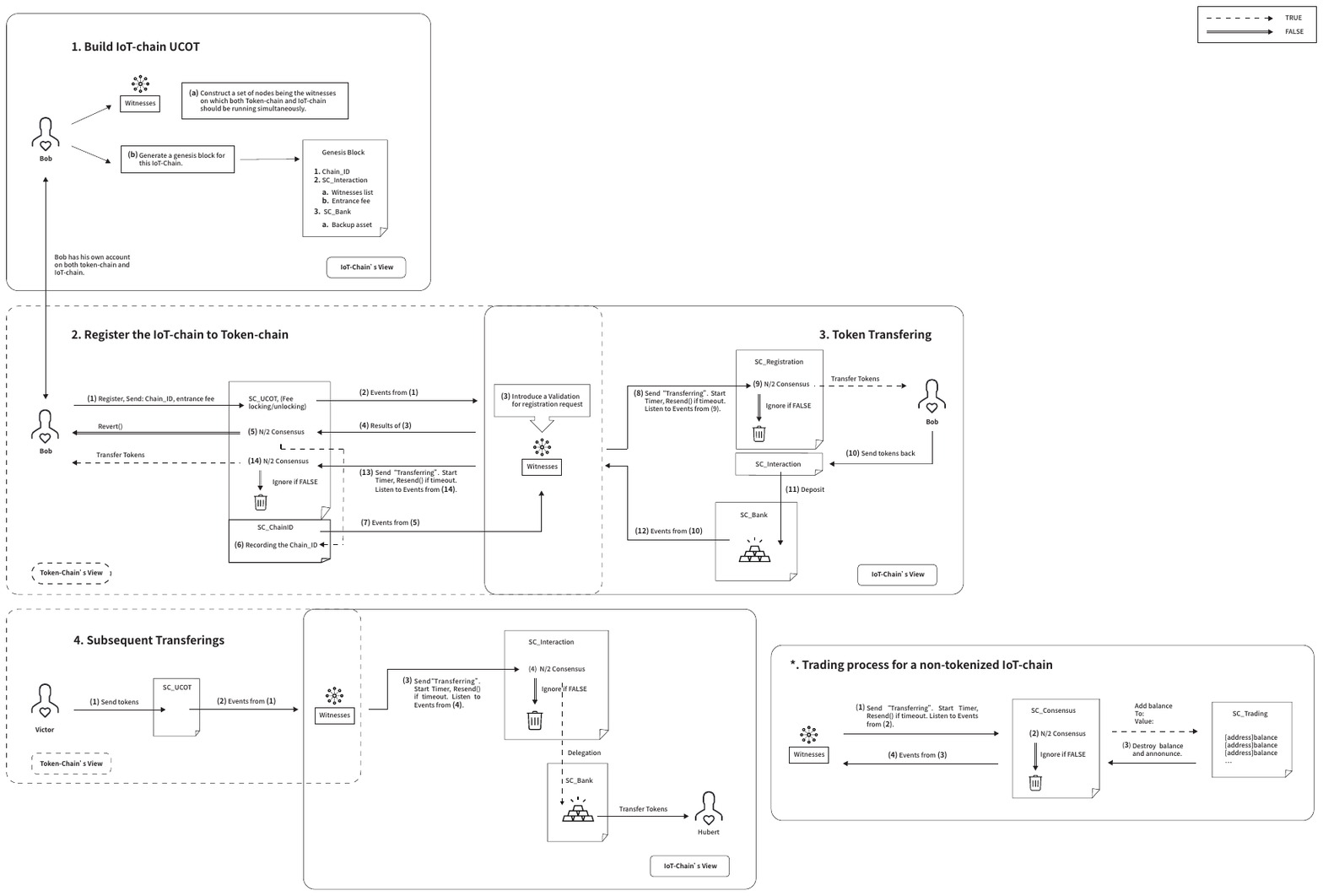}
\caption{Architecture of the invention}
\label{cross}
\end{figure}

\begin{figure}[htpb]
\centering
\includegraphics[width=5in]{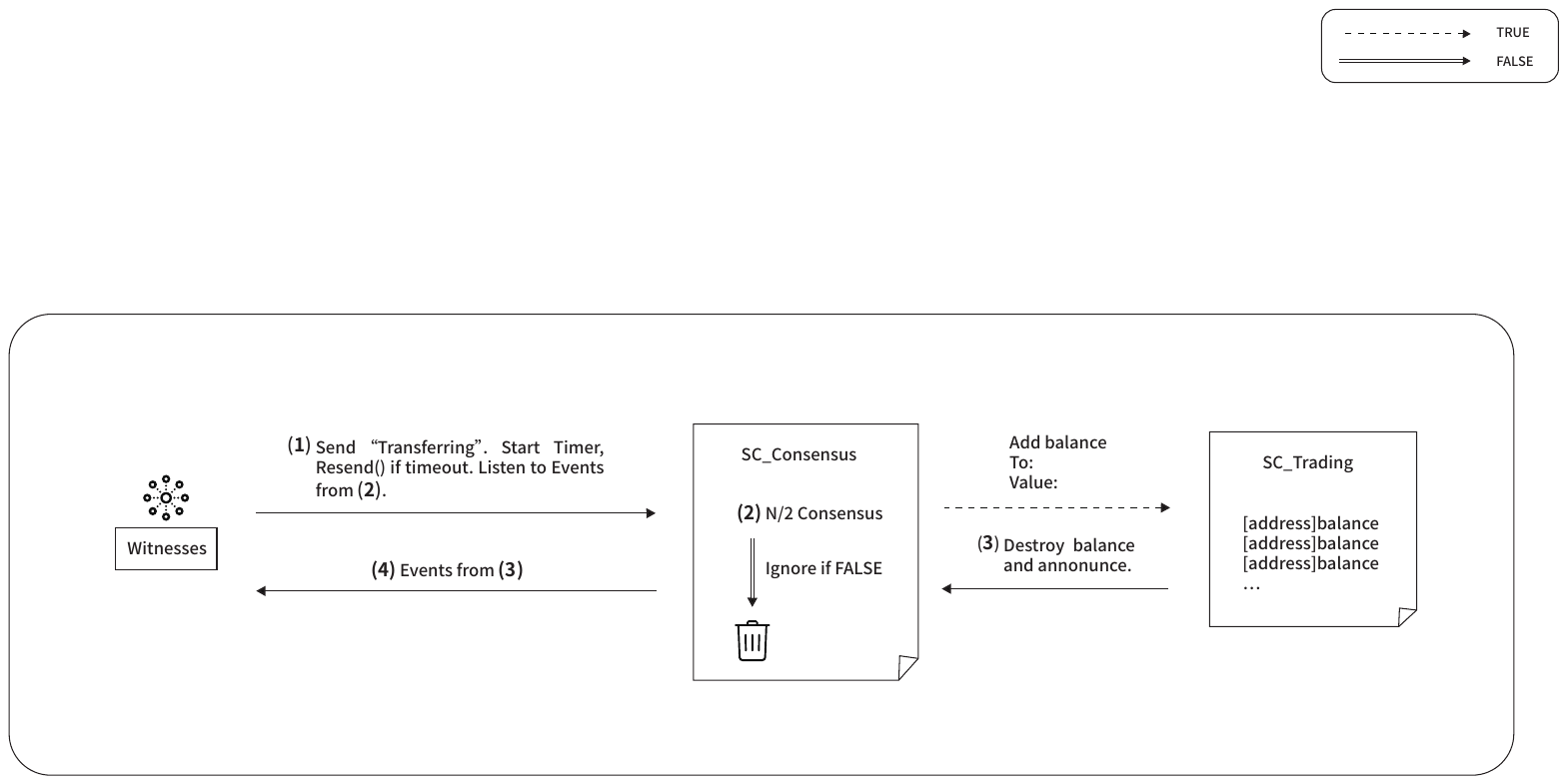}
\caption{Flow chart of trading on smart contract for a side chain with the native gas}
\label{cross_8}
\end{figure}

\begin{enumerate}
\item Figure \ref {cross} shows the flow diagram for the entire architecture of our invention. It combines the rest of the Figures into a whole to elaborate the entire flow of the procedure.
\item Figure \ref {cross_1} shows the flow diagram for the pre-defined parameters required to be setup prior to the registration to \textit{Token Chain} for an \textit{IoT Chain} without the native gas.
\item Figure \ref {cross_2} shows the flow diagram for the first time registration to \textit{Token Chain} for an \textit{IoT Chain} without the native gas from the perspective of \textit{Token Chain}.
\item Figure \ref {cross_3} shows the flow diagram for the first time registration to \textit{Token Chain} for an \textit{IoT Chain} without the native gas from the perspective of \textit{IoT Chain}.
\item Figure \ref {cross_4} shows the flow diagram for transferring consumable token from \textit{IoT Chain} back to \textit{Token Chain} from the perspective of \textit{IoT Chain}.
\item Figure \ref {cross_5} shows the flow diagram for transferring consumable token from \textit{IoT Chain} back to \textit{Token Chain} from the perspective of \textit{Token Chain}.
\item Figure \ref {cross_6} shows the flow diagram for subsequently transferring from \textit{Token Chain} to \textit{IoT Chain} that has been registered from the perspective of \textit{Token Chain}.
\item Figure \ref {cross_7} shows the flow diagram for subsequently transferring from \textit{Token Chain} to \textit{IoT Chain} that has been registered from the perspective of \textit{IoT Chain}.
\item Figure \ref {cross_8} shows the flow diagram for transferring between \textit{Token Chain} and an \textit{IoT Chain} with the native gas.
\end{enumerate}
\section{Brief Description of the Smart Contracts and Witnesses}
\begin{itemize}
\item \textit{SC\_A}: A dedicated cross-chain smart contract for an \textit{IoT Chain \{A\}}, deployed on \textit{Token Chain}. It locks/unlocks the assets transferred. It introduces the consensus for the Validation of registration from witnesses, as well as the consensus for the assets transferred back to \textit{Token Chain}.
\item \textit{SC\_ID}: It is deployed on \textit{Token Chain}. A \textit{Token Chain} can only have one \textit{SC\_ID} and it offers 2-of-2 Mulsig (Note that it can be a n-of-m Mulsig that has been claimed in the last section. For the rest of place using Mulsig, 2-of-2 is implemented in order for the simple elaboration). It records the $Chain\_ID$ of \textit{IoT Chains} that have been successfully registered. It $Updates$ the list of $Chain\_ID$ once the consensus for the Validation of registration from witnesses gets passed and emits an event of $ExistOrNot$.
\item \textit{SC\_Register}: A unique dedicated cross-chain smart contract for the first time registration, deployed on an \textit{IoT Chain}. It introduces the consensus for the $entrance fee$ transferred to the \textit{IoT Chain}. Its balance is hardcoded and will be run out after the success of registration and commits suicide afterwards by the agreement of Mulsig.
\item \textit{SC\_Inter}: A unique dedicated cross-chain smart contract for the assets transferring back to \textit{Token Chain} as well as the subsequent assets transferred to \textit{IoT Chain} with a consensus process, deployed on \textit{IoT Chain}. It dedicates the assets unlocking to \textit{SC\_Bank} after the success of consensus. It forwards the assets transferred from users' accounts to \textit{SC\_Bank} for assets locking in the case of \textit{IoT-Chain}-to-\textit{Token-Chain}.
\item \textit{SC\_Bank}: A unique dedicated cross-chain smart contract for assets locking/unlocking, deployed on \textit{IoT Chain}. The balance of \textit{SC\_Bank} that equals to the total amount of the upper bound of the potential assets trading on \textit{IoT Chain} is hardcoded in the genesis block. The corresponding amount of token is unlocked and transferred to the destination address with the delegation from \textit{SC\_Inter}. It receives and locks the assets transferred from users' accounts to \textit{SC\_Bank} for assets locking in the case of \textit{IoT-Chain}-to-\textit{Token-Chain}.
\item \textit{SC\_Consensus}: A unique dedicated cross-chain smart contract for both registration and consensus in the context of an \textit{IoT Chain} with the native gas. 
\item \textit{SC\_Trading}: A unique dedicated cross-chain smart contract that records the balance for all accounts that have ever been involved in a cross-chain transaction and dynamically maintains a ledger. Both intra-transactions and inter-transactions associated with real business must be manipulated via \textit{SC\_Trading}. In other words, the token value is generated and destroyed with \textit{SC\_Trading} running on an \textit{IoT Chain} with the native gas. The amount of consumable assets is precisely defined even though its native account-balance tends to be infinite.
\item \textit{Witnesses}: A set of nodes that bridge the information exchanged between \textit{Token Chain} and a specific \textit{IoT Chain}. Information exchanged needs to be verified by the consensus on both \textit{Token Chain} and the \textit{IoT Chain}. \textit{Witnesses} retrieve events from:
\begin{enumerate}
\item \textit{SC\_ID}: if the $Chain\_ID$ has been appended in the list;
\item \textit{SC\_A}: if some numbers of cross-chain transactions have arrived;
\item \textit{SC\_Bank}: if the assets has been successfully locked in the context of \textit{IoT-Chain}-to-\textit{Token-Chain};
\item \textit{SC\_Consensus}: if some numbers of cross-chain transactions have arrived.
\end{enumerate}

\end{itemize}
\section{Detailed Description}
A set of pre-defined \textit{Witnesses} exchange the cross chain messages as if a bridge connects \textit{Token Chain} and an \textit{IoT Chain}. Remark that multiple \textit{IoT Chains} can be implemented along with their own pre-defined \textit{Witnesses}.
All \textit{IoT Chains} are pegged to \textit{Token Chain}, which provides an individual channel for each \textit{IoT Chain} to exchange token. \textit{IoT Chains} do not interact with each other directly. 

In addition, businesses tend to customized their own \textit{IoT Chains} regardless of a specific type of consensus algorithm or even the existence of native gas, i.e. either of an \textit{IoT Chain} with native gas or an \textit{IoT Chain} without native gas. Note that the native gas refers to the native gas-oriented account-balance. Also note that the native gas-oriented account-balance is firstly implemented in Ethereum platform. It is often used to restrict the malicious transactions flooding among the network. The difference between them is shown as follows,
\begin{itemize}
\item \textbf{\textit{IoT Chain} without the native gas: }an \textit{IoT Chain} has no native gas-oriented account-balance. Nodes on such an \textit{IoT Chain} cannot send transactions without the imported gas transferred from the corresponding \textit{Token Chain}, where the imported gas, in fact, is consuming the main tokens. In other words, the main tokens are responsible for both sending a transaction and dealing with a business issue.
\item \textbf{\textit{IoT Chain} with the native gas: }an \textit{IoT Chain} has the native gas-oriented account-balance. Nodes on such an \textit{IoT Chain} can send transactions (uploading the IoT messages) without the imported gas transferred from the corresponding \textit{Token Chain}. However, any transactions associated with the business issue must come along with the imported gas.
\end{itemize}

\textit{IoT Chains} are free to be customized with low restrictions in terms of the consensus algorithm and the existence of native gas. All tokens on \textit{IoT Chains} are originated from \textit{Token Chain}, referring to the main tokens. In other words, \textit{IoT Chains} have no original token value regardless of the presence of the native gas. 
Note that \textit{IoT Chain} is allowed to have its own account-balance mechanism to maintain the whole system running fluently with a necessary access control and a flow control. This account-balance is meant to be a sufficiently large amount of value that can be automatically recharged by administrators, which implies that only the token originated from \textit{Token Chain} that is strictly differentiated from the native account-balance on \textit{IoT Chains}, has the real economical value. Thus users can choose to either registering an \textit{IoT Chain} without the native gas in which only the token originated from \textit{Token Chain} dominates the system, or an \textit{IoT Chain} with the native gas in which its own account-balance mechanism is in charge of its self-driven processing, which are both independent to the consensus algorithm.

\begin{figure}[htpb]
\centering
\includegraphics[width=5in]{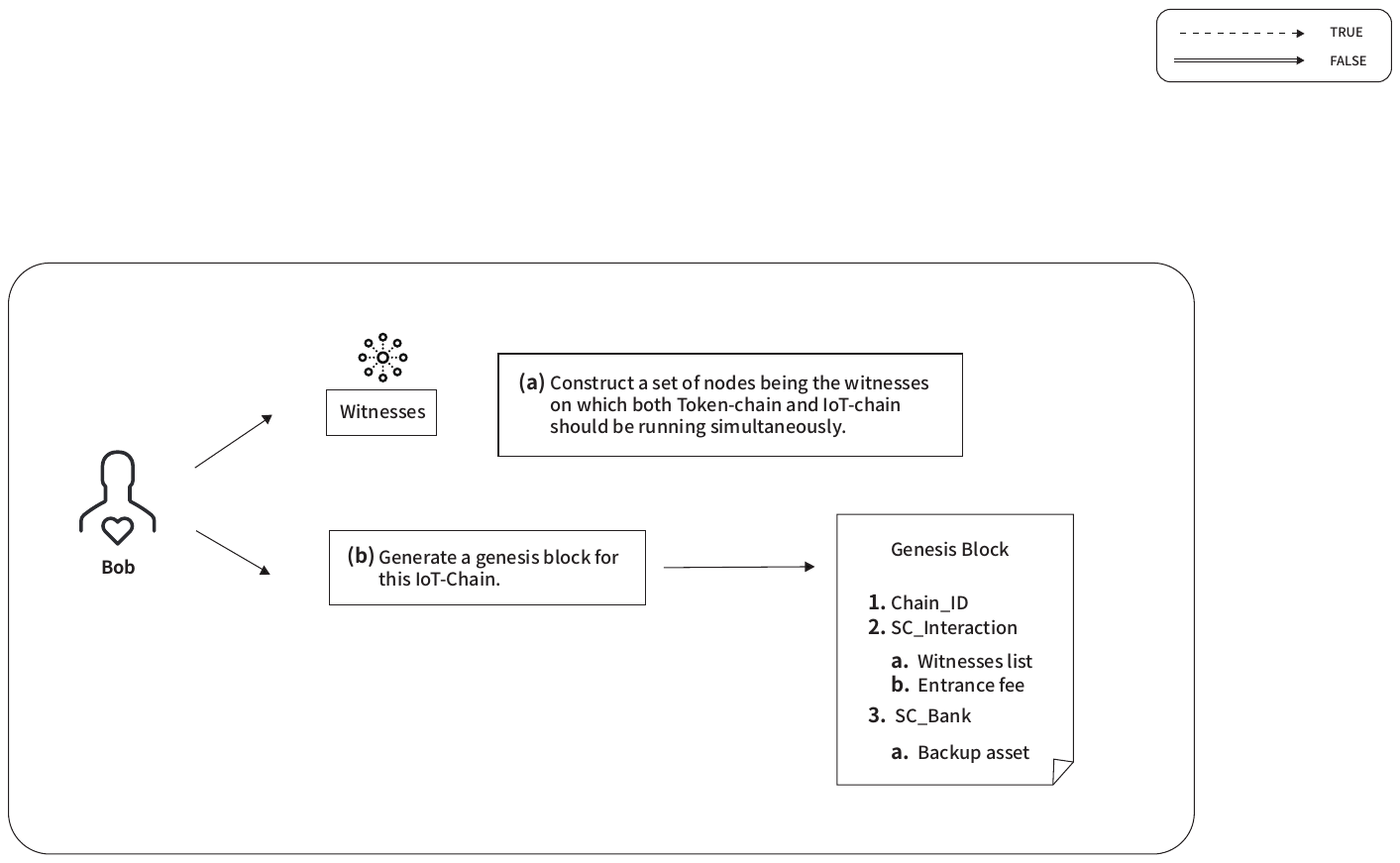}
\caption{Pre-definition}
\label{cross_1}
\end{figure}

\begin{figure}[htpb]
\centering
\includegraphics[width=5in]{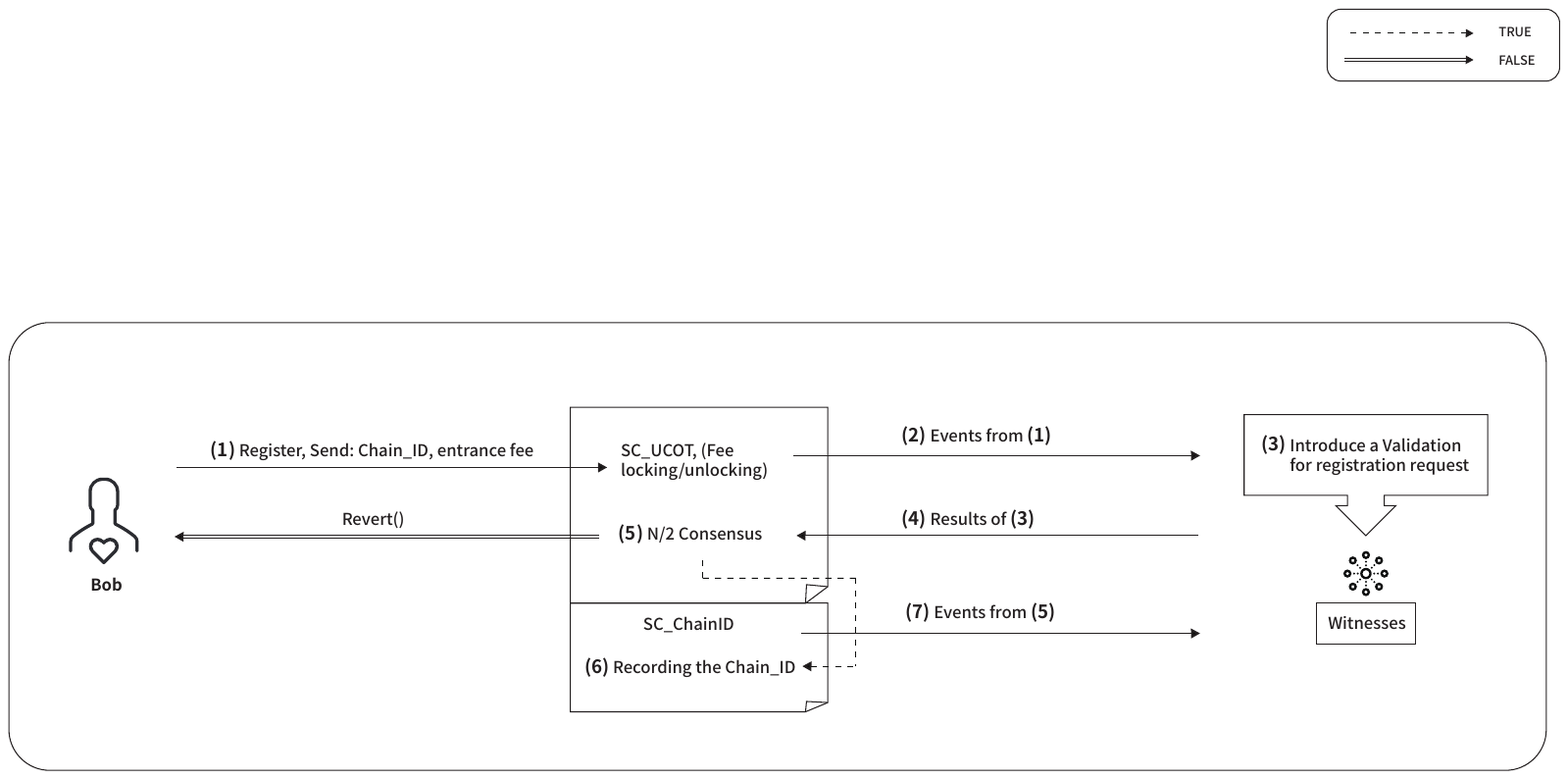}
\caption{Registration}
\label{cross_2}
\end{figure}

\begin{figure}[htpb]
\centering
\includegraphics[width=5in]{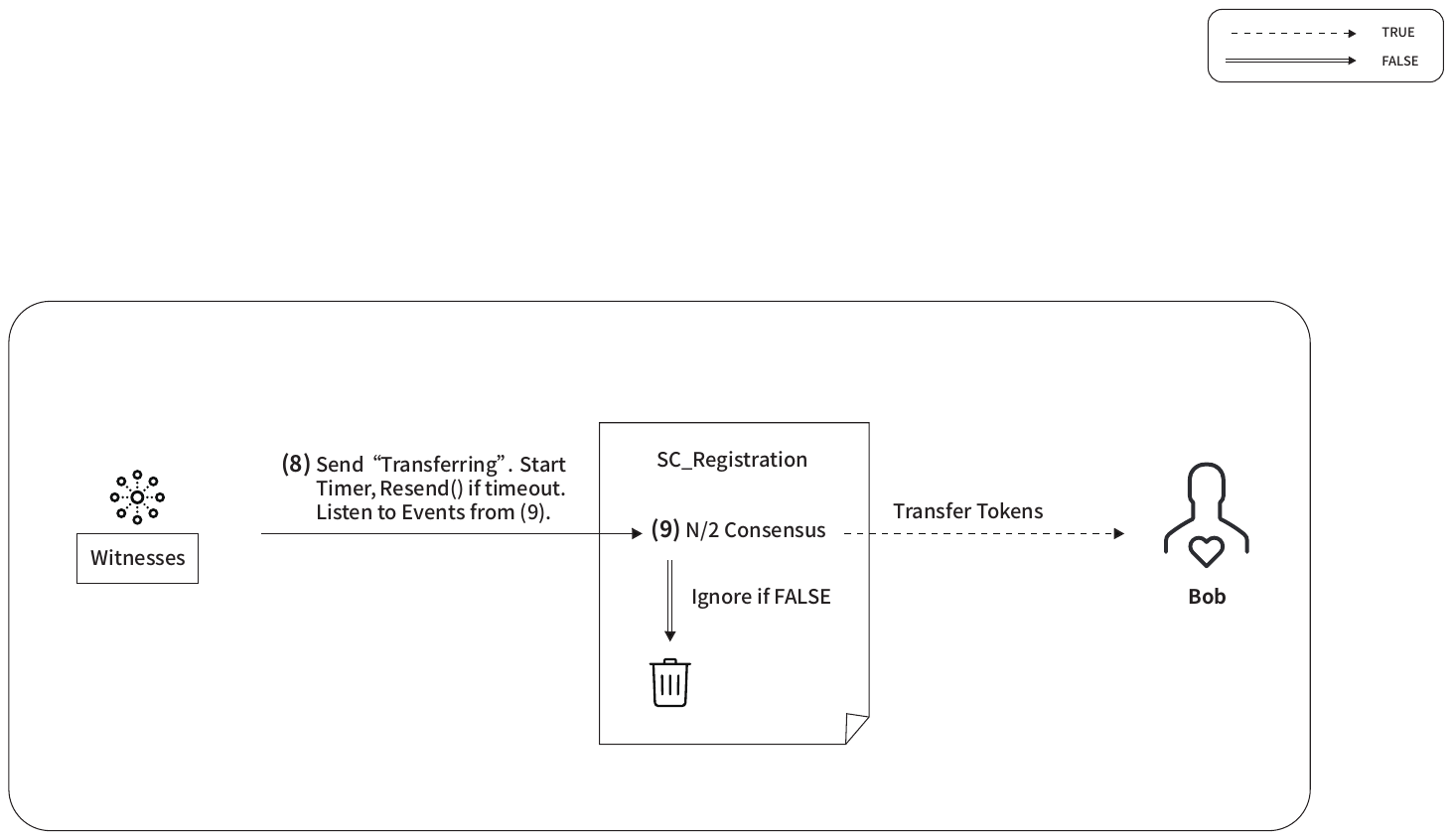}
\caption{First time assets transferring}
\label{cross_3}
\end{figure}

\begin{figure}[htpb]
\centering
\includegraphics[width=5in]{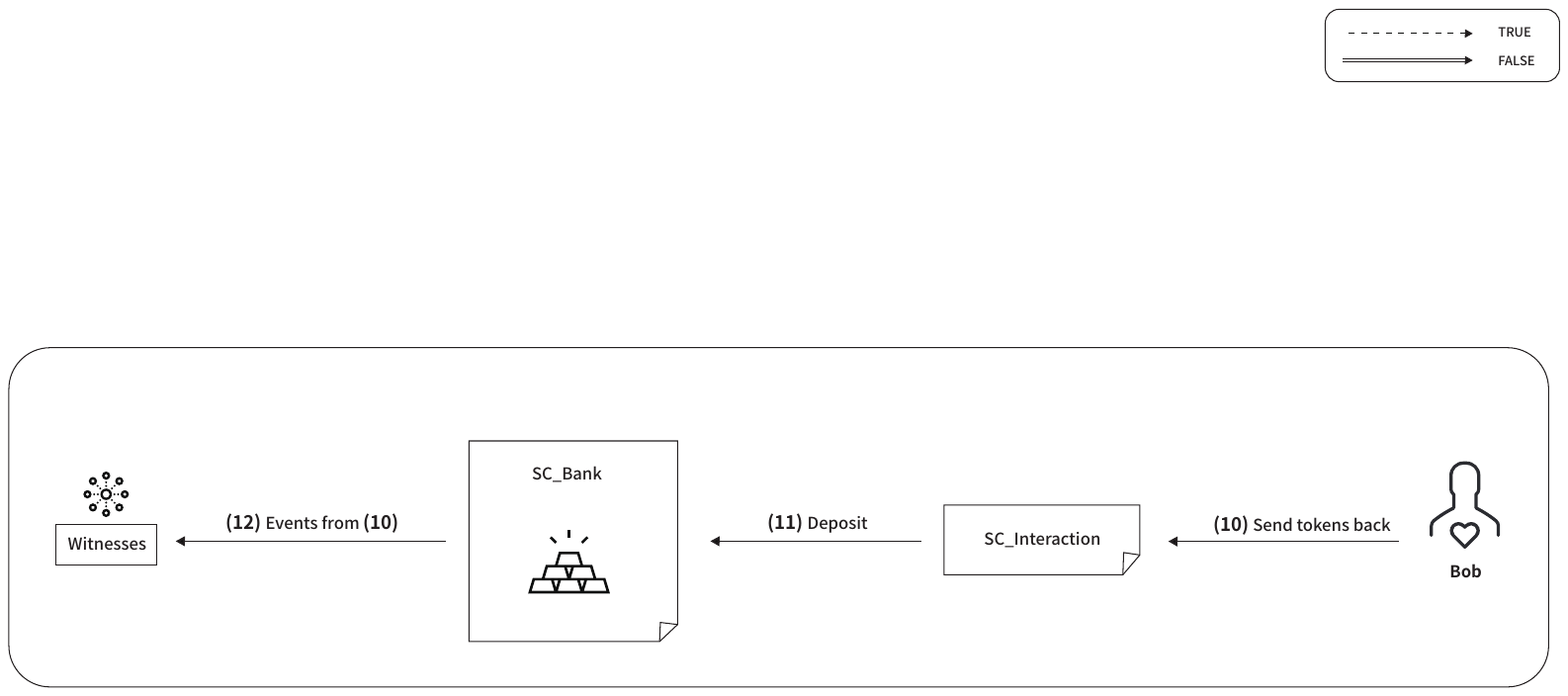}
\caption{Backward assets transferring in the view of side chain}
\label{cross_4}
\end{figure}

\begin{figure}[htpb]
\centering
\includegraphics[width=5in]{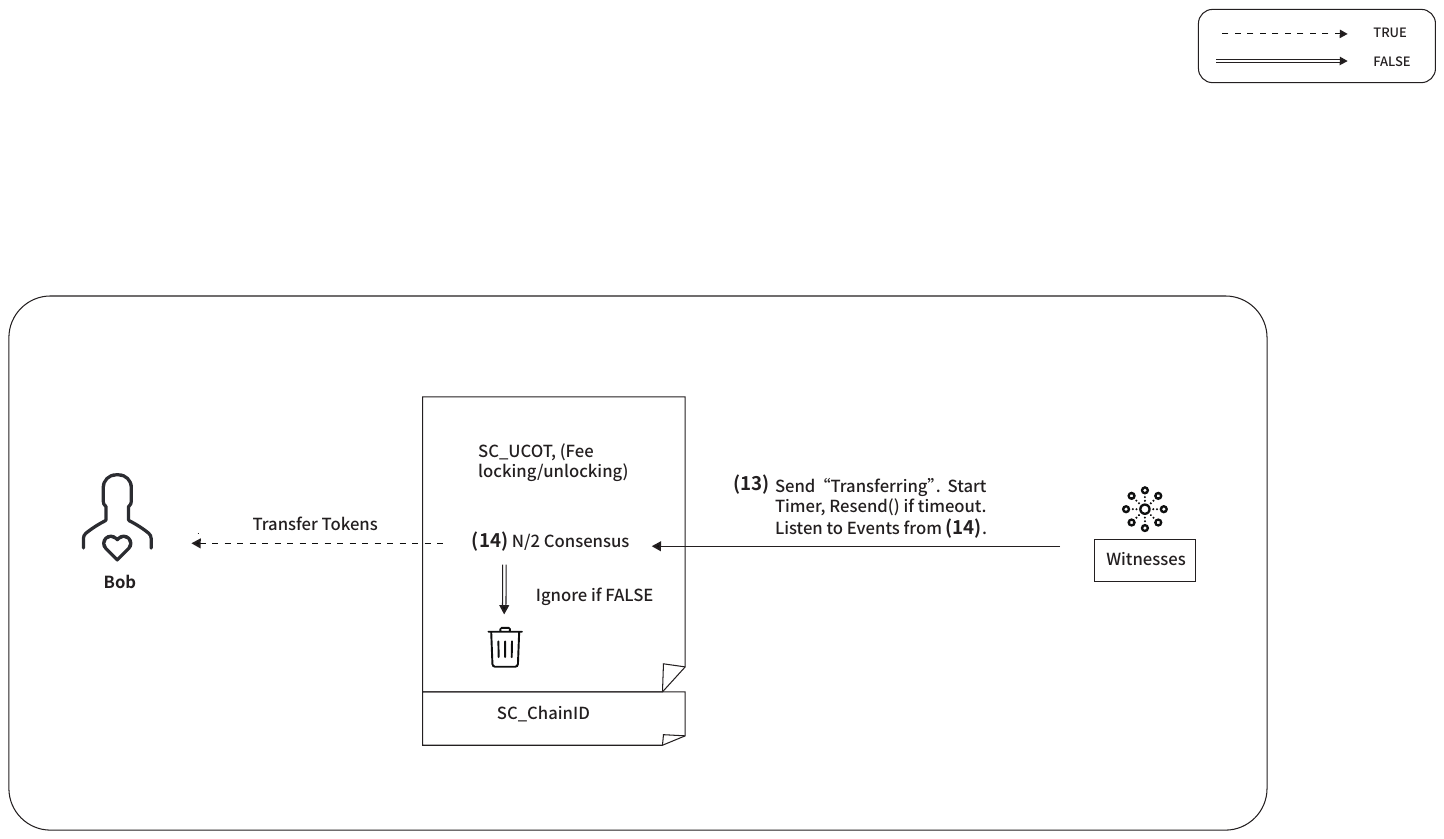}
\caption{Backward assets transferring in the view of token chain}
\label{cross_5}
\end{figure}

\begin{figure}[htpb]
\centering
\includegraphics[width=5in]{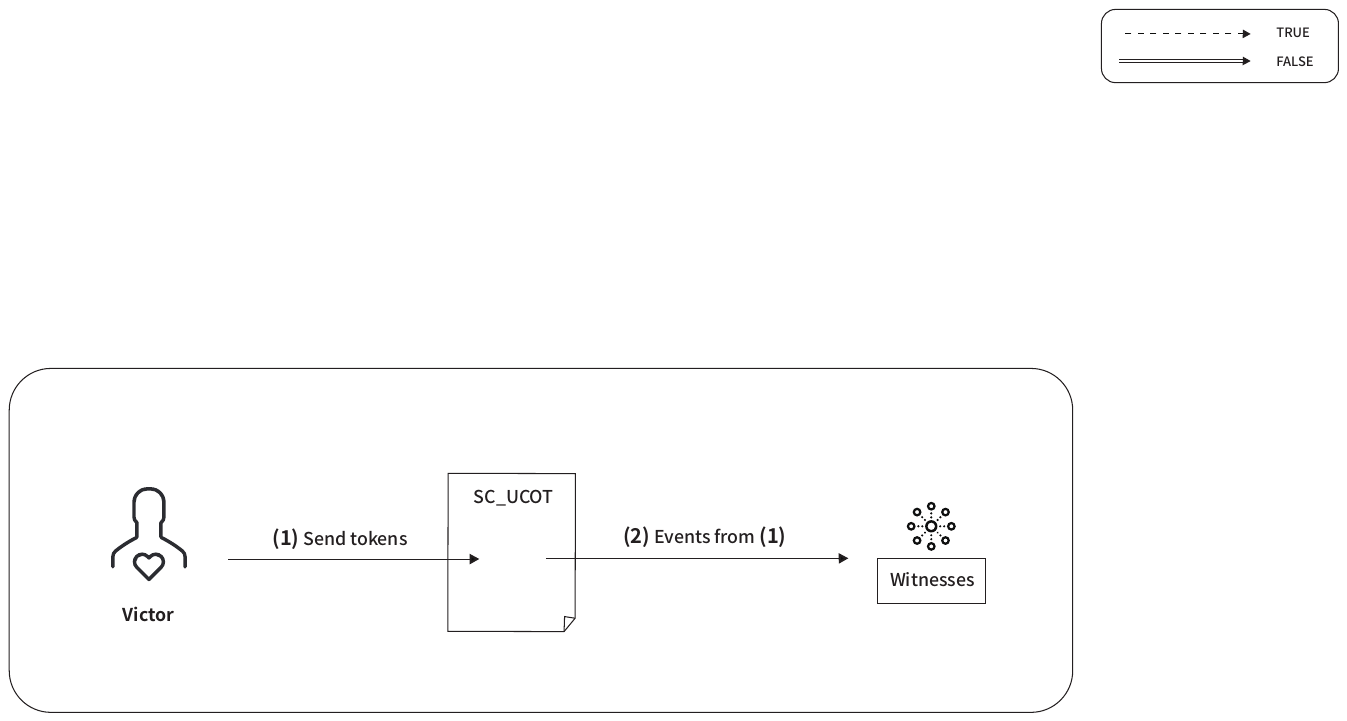}
\caption{Subsequent assets transferring in the view of token chain}
\label{cross_6}
\end{figure}

\begin{figure}[htpb]
\centering
\includegraphics[width=5in]{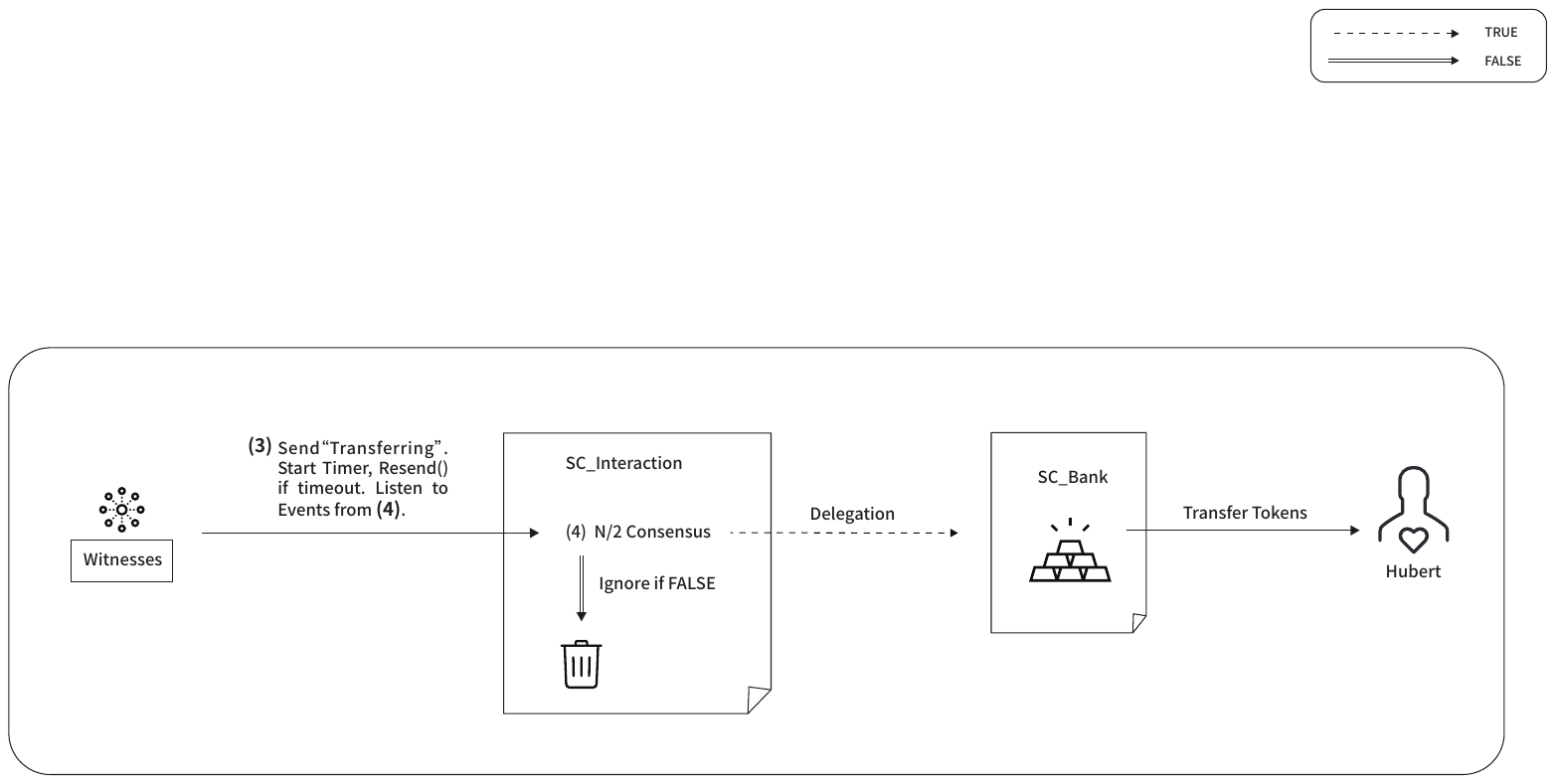}
\caption{Subsequent assets transferring in the view of side chain}
\label{cross_7}
\end{figure}

For the rest of this section we extend the description to a deeper extent shown as follows,
\begin{itemize}
\item Section 5.1 refers to Figure \ref {cross_1}, \ref {cross_2}, \ref {cross_3}, \ref {cross_4}, \ref {cross_5}, \ref {cross_6} and \ref {cross_7}.
\begin{itemize}
\item Section 5.1.1 refers to Figure \ref {cross_1}.
\item Section 5.1.2 refers to Figure \ref {cross_2} and \ref {cross_3}.
\item Section 5.1.3 refers to Figure \ref {cross_6} and \ref {cross_7}.
\item Section 5.1.4 refers to Figure \ref {cross_4} and \ref {cross_5}.
\end{itemize}
\item Section 5.2 refers to Figure \ref {cross_8};
\item Section 5.3 describes the mechanism of the reverting and resending process of our invention when the consensus fails.
\end{itemize}
Note that the entire flow chart refers to Figure \ref {cross}.

\subsection{Bridging a Side Blockchain without the native gas}
Bridging a side Blockchain without the native gas includes the following steps,
\begin{enumerate}
\item pre-definition of necessary parameters;
\item registration of the chain;
\item subsequent assets transferring:
\begin{itemize}
\item transferring from \textit{IoT Chain} back to \textit{Token Chain};
\item transferring from \textit{Token Chain} to \textit{IoT Chain} after the registration.
\end{itemize}
\end{enumerate}

\subsubsection{Pre-definition}
A set of \textit{Witnesses} run as intermediaries to bridge \textit{Token Chain} and an specific \textit{IoT Chain}. These \textit{Witnesses} have valid accounts on both \textit{Token Chain} and the \textit{IoT Chain}. The interactions between \textit{Token Chain} and the \textit{IoT Chain} are secured by Blockchain. Only interactions verified and recorded in the form of transactions are accepted and performed on both sides. Smart contracts implement the interactions between \textit{Token Chain} and the \textit{IoT Chain}. 

The genesis block of the \textit{IoT Chain} has a pre-defined form. The genesis block defines the account states and the interaction parameters of the \textit{IoT Chain}. In the genesis block, all accounts on the \textit{IoT chains} have no balance when the \textit{IoT Chain} initializes, except that two pre-defined smart contracts created by the creator of the \textit{IoT Chain} stating the balance of these smart contracts and promises to lock \textbf{\textit{some fee}} (shown below) from these addresses, respectively. Each of them is verified by the signature of its own key.
\begin{enumerate}
\item the first kind of fee is assigned as the \textit{entrance fee} that is the minimum amount of token required to initialize an interactive \textit{IoT Chain} for the registration phase. \emph{Note that the \textit{entrance fee} should be minimum-bounded to prevent from maliciously frequent registration requests};
\item the second kind of fee is assigned as the \textit{reserved assets} from which the subsequent assets can be unlocked.
\end{enumerate}
The \textit{IoT Chain} predefines $N$ \textit{Witnesses} in the genesis block of the \textit{IoT Chain}. These $N$ \textit{Witnesses} should have valid accounts on \textit{Token Chain}. 
In other words, a valid genesis block of an interactive \textit{IoT Chain} without the native gas has the following fields:
\begin{itemize}
\item $Chain\_ID$: The Chain ID of the \textit{IoT Chain};
\item $SC\_Register$: An interactive smart contract on the \textit{IoT Chain} which enables transfer function between the \textit{IoT Chain} and \textit{Token Chain}. 
\item $Bal\_Resv$: The amount of token that the creator reserves in $SC\_Register$ for the \textit{IoT Chain};
\item $SC\_Inter$: A smart contract on the \textit{IoT Chain},  used for subsequent transferring from \textit{Token Chain} to the \textit{IoT Chain} and transferring from the \textit{IoT Chain} back to \textit{Token Chain};
\item $SC\_Bank$: A smart contract on the \textit{IoT Chain} which enables the subsequent token locking and unlocking. There is a strict access control upon this smart contract, i.e. the only address who can interact with $SC\_Bank$ is the address of $SC\_Inter$. 
\item $Bal\_Bank$: The amount of token that the creator reserves in $SC\_Bank$ for the \textit{IoT Chain}; Note that $Bal\_Resv + Bal\_Bank \equiv$ total amount of token on \textit{Token Chain} including those unmined token;
\item $Wit\_Addr\_List$: The addresses of \textit{Witnesses} for the \textit{IoT Chain}, which is hardcoded in both $SC\_Register$ and $SC\_Inter$.
\end{itemize}

The creator creates and broadcasts the genesis block in the IoT network and await the other nodes being launched and ready for the the new \textit{IoT Chain} within an expected period. Note that any transactions with a source address of one of the \textit{Witnesses} and a destination address of \textit{SC\_Register} or \textit{SC\_Inter} can pay a zero gasprice. Meanwhile, any transactions with a source address of either \textit{SC\_Register}, \textit{SC\_Inter} or \textit{SC\_Bank} can also pay a zero gasprice.

\subsubsection{Registration}
After all \textit{Witnesses} have been ready (All \textit{Witnesses} should then be able to send transactions to and retrieving events from any targeted addresses. Meanwhile, they all are able to fetch the json files of both of the genesis blocks of the \textit{IoT Chain} and \textit{Token Chain}), the creator on \textit{Token Chain} sends a request of registration in the form of a transaction to \textit{SC\_A}, where the following contents are included in the invoked event,
\begin{itemize}
\item all token that is contained in this transaction sent to \textit{SC\_A} subtracting the compensation fee for both \textit{SC\_A} and \textit{Witnesses} providing the transferring service. It guarantees that \textit{SC\_A} and \textit{Witnesses} have sufficient fund to send transactions on \textit{Token Chain};
\item $Chain\_ID$ of the \textit{IoT Chain} that the creator proposes.
\end{itemize}
\textit{Note that \textit{SC\_A} is a smart contract deployed on \textit{Token Chain} with no balance initially, which is a dedicated cross-chain service relay and created by the service provider in the form of \textbf{open source}.} 
As such the process of validation on \textit{Witnesses} gets called by retrieving the events invoked by this requesting transaction. 
It validates the genesis block according to Alg. \ref{alg: verify IoT chain} from which the output is sent to \textit{SC\_A}. 
In other words, once a \textit{Witness} validates the genesis block, it accepts the registration request of the new \textit{IoT Chain} and sends a ``Confirm'' transaction to the smart contract, \textit{SC\_A}, on \textit{Token Chain}. 
When \textit{SC\_A} receives ``Confirm'' transactions from more than $N/2$ \textit{Witnesses} (\textit{Note that for any smart contracts on which a consensus process is being conducted, a parameter $round$ must be used to prevent from out-of-order message. 
The minimum faulty tolerance is $N/2$}), the smart contract \textit{SC\_A} reserves $Bal\_Resv$ on \textit{Token Chain} from the account of the creator, triggers the smart contract \textit{SC\_ID} to register the \textit{IoT Chain} with its $Chain\_ID$. 
Note that Alg. \ref{alg: verify IoT chain} has ensured that the $Chain\_ID$ for each chain is unique in the entire system. 
A signal indicating whether the registration has succeeded or not is contained in the event invoked by $Chain\_ID$ that \textit{Witnesses} are listening to. 
If and only if a ``True'' is received, a \textit{Witness} sends a ``Transferring'' transaction to the smart contract $SC\_Register$ on the \textit{IoT Chain}. 
If $SC\_Register$ receives ``Transferring'' transaction from more than $N/2$ \textit{Witnesses}, the same amount (this amount should now exactly equal to the total balance of $SC\_Register$, ensuring that the balance of $SC\_Register$ will be zero after unlocking) of token are unlocked from $SC\_Register$ on \textit{IoT Chain}. 

$SC\_Register$ commits suicide by the owners after the owners both (2-of-2 Mulsigs) have retrieved the event of success of adding the $Chain\_ID$ in $SC\_ID$.

After the \textit{IoT Chain} is created and registered, the \textit{IoT Chain} will be able to work properly and generate blocks based on its consensus protocol. \emph{Note that there is no block rewards on the \textit{IoT Chain}}. 

\subsubsection{Token Chain to IoT Chain after Registration}
$SC\_Inter$ is responsible for any subsequent ``Transferring'' transactions sent from \textit{Token Chain} to a specific \textit{IoT Chain}, plus those sent from the \textit{IoT Chain} back to \textit{Token Chain}. If a subsequent ``Transferring'' transaction sent to $SC\_Inter$ is detected, \textbf{$SC\_Inter$ delegates the burden to $SC\_Bank$ along with $to$ and $value$ of each cross-chain transaction after the success of $N/2$ consensus}. This process succeeds if and only if it passes the validation that the current balance of $SC\_Bank$ + the \textit{entrance fee} + the amount of token that is requested for this specific ``Transferring'' transaction $\equiv$ total amount of token on \textit{Token Chain} including those unmined token. The owners of $SC\_Bank$ add the address of $SC\_Inter$ in $SC\_Bank$ and the corresponding amount of token can be unlocked if and only if both owners agree. \emph{Note that it is none of business of $SC\_ID$ for a subsequent cross-chain transferring. \textit{Witnesses} do not listen to the events invoked by $SC\_ID$. Instead, events of $SC\_A$ are listened.}

\subsubsection{IoT Chain to Token Chain after Registration}
Suppose the aforementioned \textit{IoT Chain} is still discussed, there are two types of transactions on the \textit{IoT Chain}: one records the IoT events and does not interact with \textit{Token Chain}; the other one records the interactions with \textit{Token Chain}. Any transaction that is associated with $SC\_Inter$ belongs to Interaction transactions. 
Suppose that a user on the \textit{IoT Chain} sends a request of transferring token back to \textit{Token Chain} in the form of a transaction sent to $SC\_Inter$ with an arbitrary amount of token $\le$ (total amount of token on the \textit{Token Chain} including those unmined token $-$ the balance of $SC\_Bank$). $SC\_Inter$ then becomes the relay and transfers the token to $SC\_Bank$ for the further locking. Each of the
\textit{Witnesses} retrieves the events and sends a ``Transferring'' transaction to $SC\_A$ on \textit{Token Chain}. If $SC\_A$ receives a ``Transferring'' transaction from more than $N/2$ \textit{Witnesses}, the same amount (this amount should $\le$ the balance of $SC\_A$) of token is unlocked from $SC\_A$ on \textit{Token Chain}. 

\begin{algorithm}  
\caption{Verify the Request of Registering an IoT Chain}  \label{alg: verify IoT chain}
\LinesNumbered  
\KwIn{

$\quad$$Hash(Genesis)$: the hash of the IoT genesis block. \\
$\quad$$Balance_{request}$: the total balance that the creator reserves for the IoT chain. \\
$\quad$$Balance_{json}$: the total balance that the json file of the IoT chain is showing.\\
$\quad$ $ChainID_{request}$: the chain ID that the creator proposes for the IoT chain. \\
$\quad$ $ChainID_{json}$: the chain ID that the json file of the IoT chain is showing.\\
 $\quad$ $ChainID_{token}$: the chain ID of the token chain.\\
$\quad$ $SmartContract_{ChainID}$: a list of all chain IDs that have been successfully registered already.} 
\KwOut{True or False}
\textbf{The witnesses assert the following conditions:}

Confirm the integrity of the Genesis Block: 

$\quad$$GetHash(Genesis)$ = $Hash(Genesis)$;

Confirm it is a request to generate a new IoT Chain:

$\quad$$GetHeight()$ = 0;

Verify the balance of the creator: 

$\quad$$Balance_{request}$ = $Balance_{json}$

Validate the Chain ID:

$\quad$$ChainID_{request}$ = $ChainID_{json}$

$\quad$$ChainID_{request}$ $\neq$ $ChainID_{token}$

$\quad$$ChainID_{request}$ $\notin$ $SmartContract_{ChainID}$

If ALL above conditions are satisfied, the output is ``TRUE'', which indicates the witness validates the registration request; or else the output is ``FALSE'', which indicates the witness refuses the registration request. 
\end{algorithm}

\subsection{Bridging a Side Blockchain with the native gas}
Comparing with bridging an \textit{IoT Chain} without the native gas, there are only a few tiny changes to bridge an \textit{IoT Chain} with the native gas, which is shown as follows,
\begin{enumerate}
\item \textit{SC\_Register} and \textit{SC\_Inter} merge into one smart contract, \textit{SC\_Consensus}, since there is no need to transfer token itself. Instead, the only thing needs to do is to update the balance in \textit{SC\_Trading}. \textit{SC\_Consensus} has two hardcoded owners, offering 2-of-2 MultiSig;
\item A set of \textit{Witnesses} is still needed to be organized and written in \textit{SC\_Consensus}. Note that there is no need to predefine \textit{SC\_Consensus} in the genesis block as long as it provides sufficient native account-balance for self-driven processing. Also, there is no need to set an \textit{entrance fee} in \textit{SC\_Consensus};
\item \textit{SC\_Bank} can be deprecated;
\item It is still required to have a smart contract running on \textit{Token Chain}, \textit{SC\_A} created by the service provider in the form of \textbf{open source}. However, the events fetched by \textit{Witnesses} for the Validation only contain the information of how much is transferred and the \textit{Chain\_ID}. Accordingly, \textit{Witnesses} only validate the \textit{Chain\_ID} part in \textit{Algorithm 1} regardless of the height and balance of the \textit{IoT Chain};
\item As if a trading platform manipulate the financial trading, \textit{SC\_Trading} records the balance for all accounts that have ever been involved in an arbitrary cross-chain transaction and dynamically maintains a ledger based on the historical balance. Both intra-transactions and inter-transactions associated with real business must be manipulated via \textit{SC\_Trading}. In other words, the token value is generated and destroyed with \textit{SC\_Trading} running on an \textit{IoT Chain} with the native gas. In the context of high throughput of the \textit{IoT Chain} along with its native account-balance mechanism, it is believed that \textit{SC\_Trading} is able to handle the transactions in a high volume. Note that the following statement must be satisfied,
$$\left(\sum_{i=0}^{N}Bal_i\right)_{ChainID}=Asset_{ChainID},$$
where the total amount of token that is being locked in \textit{SC\_A} with a certain $Chain\_ID$ must equal to the sum of the balance of all accounts recorded in \textit{SC\_Trading} on an \textit{IoT Chain} with the same $Chain\_ID$. It has two hardcoded owners, offering 2-of-2 MultiSig.
\item Either of \textit{SC\_Consensus} or \textit{SC\_Trading} can be pre-defined in the genesis block without having the owners and offering the Mulsig if the r \textit{IoT Chain} with the native gas is at the height of zero when initializing. As such a valid genesis block of an \textit{IoT Chain} with the native gas has the following fields:
\begin{itemize}
\item $Chain\_ID$: The Chain ID of the \textit{IoT Chain};
\item $SC\_Register$: An interactive smart contract on the \textit{IoT Chain} which enables transfer function between the \textit{IoT Chain} and the \textit{Token Chain}; 
\item $SC\_Trading$: An smart contract on the \textit{IoT Chain} which records the balance for all accounts that have ever been involved in an arbitrary cross-chain transaction and dynamically maintains a ledger based on the historical balance;
\item $Wit\_Addr\_List$: The addresses of \textit{Witnesses} for the \textit{IoT Chain}, which is hardcoded in both $SC\_Register$ and $SC\_Inter$.
\end{itemize}

\end{enumerate}

\subsection{Process of Reverting and Resending}
Note that there exist multiple solutions to the failure of consensus, such as $Revert$ or $Resend$. In this documentation, we also provide our mechanism shown as follows.
\begin{itemize}
\item \textbf{Registration:} if the consensus of $N/2$ on $SC\_A$ fails to be reached within a period of $t$ (can instead use block height $h$), the token is reverted to the account of the creator on \textit{Token Chain}, with a certain amount of compensation fee paid to $SC\_A$ and \textit{Witnesses}.
\item \textbf{Transferring:} suppose that $h_l$ denotes the height of the current latest block and $\omega$ denotes the unconfirmation window size (a public chain needs such a window size to determine the sufficiently high confidence interval of a block not being reverted by the chain-reorganization). If it is the first time of transferring, each of \textit{Witnesses} waits until $h_l-2\omega+1 \geq 0$, and collects all events associated with the transactions whose destination address is $SC\_A$, $SC\_Register$ or $SC\_Inter$ from block of $h_l-2\omega+1$ to $h_l-\omega$, which is assigned as the data of a ``Transferring'' transaction for the following consensus. If the consensus of $N/2$ in $SC\_Register$, $SC\_Inter$ or $SC\_A$ fails to be reached within a period of $t$ (can instead use block height $h$), each of \textit{Witnesses} sleeps for $T$ (now we let $h_{l2}$ denotes the latest block) and afterwards starts to resend the blocks getting stuck with the height from $h_l-2\omega+1
$ to $h_l-\omega$ (totally $\omega$ blocks). This process of resending is iterated until the consensus of $N/2$ is reached ($h_{l2}$ updates per iteration). Note that a process of resending indicates that it is not the first time of transferring. In this case, each of \textit{Witnesses} finds its $h_{l2}$ and checks whether $h_{l2} \geq h_l-\omega + 2\omega$. If not, each of \textit{Witnesses} sleeps for $T$ and starts over. Otherwise, each of \textit{Witnesses} collects the corresponding events from the blocks with height from $h_l-\omega+1$ to $h_l$ and sends the cross-chain message to $SC\_Register$, $SC\_Inter$ or $SC\_A$ for the following consensus. Also note that each of \textit{Witnesses} listens to the event invoked by the corresponding consensus of $N/2$ in order for the awareness of the result.
\end{itemize}

\section{Claims}
What is claimed is:
\begin{enumerate}
\item A method for transferring assets from a parent chain to multiple side chains and backwards, the method comprising:
\begin{itemize}
\item pre-defining, for a new side chain without the native gas, a set of parameters must be set prior to every other step in order for the later validation;
\item registering, via the Validation and Consensus by \textit{Witnesses}, the information of the side chain gets recorded by an open smart contract on the parent chain;
\item first time sending, which represents the first time transferring from the parent chain to the side chain. A certain amount of token being locked in another open smart contract on the parent chain and the same amount of token is unlocked and released by an open smart contract on the side chain;
\item reversely sending, which represents the transferring from the side chain back to the parent chain. A certain amount of token is locked on the side chain, and the same amount of token is unlocked on the parent chain if the consensus is passes;
\item subsequent sending, which represents the subsequent transferring from the parent chain to the side chain. It differs from the first time transferring in that,
\begin{itemize}
\item \textit{Witnesses} listen to the events from the smart contract that locks the assets but not the smart contract that records the Chain ID of side chains;
\item the smart contract that releases and unlocks the assets on the side chain in the first time transferring now instead delegates this to an open smart contract that locks all the rest of the potential assets.
\end{itemize}
\item reverting and resending, by the consensus whose result \textit{Witnesses} keeps listening to. Revert is implemented in the consensus of registration while Resend is implemented in the other consensuses;
\item trading, by an open smart contract on a new side chain with the native gas. All transactions associated with cross-chain require to be manipulated by this smart contract.
\end{itemize}
\item The method of claim \textbf{1}, wherein users can choose to either registering a side Blockchain without the native gas in which only the token originated from \textit{Token Chain} dominates the system, or a side Blockchain with the native gas in which its own account-balance mechanism is in charge of its self-driven processing. A side Blockchain without the native gas must be at a height of zero when registering, while a side Blockchain without the native gas can be either a new one or an existing one.
\item The method of claim \textbf{1}, wherein the MulSig is not limited to 2-of-2, can be n-of-m instead.
\item The method of claim \textbf{1}, wherein the Mulsig of \textit{SC\_Register} is for suicide. The Mulsig of \textit{SC\_Bank} is for adding the address of \textit{SC\_Inter} and subsequent assets locking and unlocking. 
\item The method of claim \textbf{1}, wherein either \textit{SC\_Consensus} or \textit{SC\_Trading} can be pre-defined in the genesis block instead of offering Mulsig if the side chain with the native gas is at the height of zero.
\item The method of claim \textbf{1}, wherein the native gas is not limited to Ethereum platform. All Ethereum-based platforms or other platforms that are self-driven by some kinds of native account-balance are also compatible.
\item The method of claim \textbf{1}, wherein the faulty tolerance of consensus can be set to any number greater than \textit{N/2}.
\item The method of claim \textbf{1}, wherein the \textit{entrance fee} is minimum-bounded to prevent from maliciously frequent registration requests.
\item The method of claim \textbf{1}, wherein the format of data by which \textit{Witnesses} validate the validity of a new side chain is not limited to json file.
\item The method of claim \textbf{1}, wherein there is no need to pay fee for block generator for any cross-chain-related smart contracts and each of \textit{Witnesses} sending a transaction with a destination address of a cross-chain-related smart contract.
\item The method of claim \textbf{1}, wherein the cross-chain-related smart contract initializes with no balance on the parent chain.
\item The method of claim \textbf{1}, wherein for any smart contracts on which a consensus process is being conducted, a parameter \textit{round} is used to prevent from out-of-order messages.
\item The method of claim \textbf{1}, wherein there is no block rewards on the side chain.
\end{enumerate}


\bibliographystyle{IEEEtran}
\bibliography{IEEEabrv,bib}

\clearpage
\section*{Appendix}

\end{document}